\documentclass[aip,amsmath,amssymb,reprint,floatfix,10pt, onecolumn]{revtex4-2}

\usepackage[english]{babel}

\usepackage{dcolumn}
\usepackage{bm}
\usepackage{subcaption}
\usepackage{xcolor}
\usepackage{times}
\usepackage{amssymb,amsmath}
\usepackage{graphicx}
\usepackage{natbib}
\usepackage{multirow}
\usepackage{float}
\usepackage{braket}

\begin{document}

\title{Magnetic Dissipation in Ferrofluids}

\author{Lili Vajtai}
\affiliation{{Stavropoulos Center for Complex Quantum Matter, Department of Physics and Astronomy, University of Notre Dame, Notre Dame, Indiana 46556, USA}}
\affiliation{{Department of Physics, Institute of Physics, Budapest University of Technology and Economics, Műegyetem rkp. 3., H-1111 Budapest, Hungary}}

\author{Norbert~M.~Nemes}
\affiliation{GFMC, Departamento de Fisica de Materiales Universidad Complutense de Madrid, 28040}
\affiliation{Instituto de Ciencia de Materiales de Madrid, 28049 Madrid, Spain}

\author{Maria del Puerto Morales}
\affiliation{Instituto de Ciencia de Materiales de Madrid, 28049 Madrid, Spain}
\affiliation{GFMC, Unidad Asociada ICMM-CSIC "Laboratorio de Heteroestructuras con Aplicaci\'on en Espintronica", Departamento de Fisica de Materiales Universidad Complutense de Madrid, 28040}

\author{Bence~G.~M\'{a}rkus}
\affiliation{{Stavropoulos Center for Complex Quantum Matter, Department of Physics and Astronomy, University of Notre Dame, Notre Dame, Indiana 46556, USA}}

\author{L\'{a}szl\'{o}~Forr\'{o}}
\affiliation{{Stavropoulos Center for Complex Quantum Matter, Department of Physics and Astronomy, University of Notre Dame, Notre Dame, Indiana 46556, USA}}

\author{Ferenc~Simon\email{simon.ferenc@ttk.bme.hu}}
\affiliation{{Stavropoulos Center for Complex Quantum Matter, Department of Physics and Astronomy, University of Notre Dame, Notre Dame, Indiana 46556, USA}}
\affiliation{{Department of Physics, Institute of Physics, Budapest University of Technology and Economics, Műegyetem rkp. 3., H-1111 Budapest, Hungary}}
\affiliation{{Institute for Solid State Physics and Optics, HUN-REN Wigner Research Centre for Physics, PO. Box 49, H-1525, Hungary}}

\date{\today}

\begin{abstract}
    Ferrofluids, composed of magnetic nanoparticles suspended in a non-magnetic carrier liquid, have attracted considerable attention since their discovery in the 1960s. Their combination of liquid and magnetic properties gives rise to complex behaviors and unique functionalities, enabling a wide range of technological applications. Among these is the ability of the magnetic material to be moved by and to absorb heat when exposed to an external magnetic field -- a process that can occur through various dissipation mechanisms depending on the system. A detailed understanding of these mechanisms is crucial for tailoring materials to specific applications. We provide a comprehensive overview of the theoretical principles underlying different energy dissipation processes and propose a coherent framework for their interpretation. Particular attention is devoted to describing the frequency-dependent susceptibility, which is the key parameter to describe dissipation. We demonstrate that dissipation, predicted from magnetometry-based studies, matches well with direct, frequency-dependent calorimetric results, expanding the available frequency range of the characterization. The demonstrating measurements were carried out with a dilute ferrofluid containing magnetite nanoparticles of a mean diameter of 10.6 nm.
\end{abstract}

\maketitle

\section{Introduction}

Ferrofluids are a fascinating class of magnetic materials that simultaneously exhibit magnetic and fluidic properties. They are composed of magnetic nanoparticles of various materials, shapes, and sizes, dispersed in a nonmagnetic carrier liquid. This structure gives rise to many complex phenomena, as a result of hydrodynamics, magnetic properties, and the additional degrees of freedom of rotation and translation of the particles within the medium \cite{Philip2023, Kole2021}. 

These unique properties led to various technological applications. Ferrofluids and nanoparticles in general are currently used as part of mechanical systems, such as magnetically controllable lubricants \cite{torresdiaz2014, Yang2022}, seals \cite{Philip2023, Rosensweig1982, torresdiaz2014, Kole2021}, cooling agents \cite{Philip2023, Rosensweig1982} or components of various sensors \cite{Baumgartner2013, Philip2023, torresdiaz2014, Yang2022}. Moreover, there is a growing interest in their use for biomedical applications, particularly as magnetic imaging agents \cite{Baumgartner2013, mri_ferrof, Shokrollahi2013, Kuznetsov2022}, in magnetically guided thermotherapy \cite{Baumgartner2013, mri_ferrof, Philip2023, Shokrollahi2013, Review_ortega_pankhurst, Hajalilou2021, Yang2022} or theranostics, i.e., where a combination of both is achieved \cite{zhou2018nanoparticles, Zhu01012017}.

Several key applications of ferrofluids are specifically engineered to maximize energy dissipation. As magnetically induced dissipation is a controllable and efficient mechanism, it is essential to understand the magnetism of the related materials. Although the ability to dissipate heat in an external magnetic field is a common feature of ferromagnetic materials, the underlying microscopic mechanisms vary with the nature of the excitation. In ferrofluids, the additional complexity of particle dynamics enables even more complicated dissipation phenomena \cite{Review_ortega_pankhurst, garaio, Review_Wu}.

Although the topic is of considerable significance, the literature often falls short in providing a thorough understanding of the distinct dissipation processes involved. In particular, dynamic mechanisms are frequently excluded from the discussion of heat loss in nanoparticle systems, even when they have a dominant contribution. This ambiguity can hinder the accurate interpretation of the underlying mechanisms and characteristics of dissipation in ferrofluids. Establishing a clear and systematic framework is crucial for advancing our understanding of these magnetic systems and enabling their precise optimization in targeted medical and technological applications. In particular, the influence of the irradiation field frequency and the dynamic response of the magnetic material -- namely, the frequency-dependent susceptibility -- are often inadequately described in the literature. Furthermore, to the best of our knowledge, limited effort has been made to correlate the magnetic properties obtained from magnetometry with calorimetric measurements \cite{WELLS2022169992,GARAIO2014432}. In calorimetry, energy dissipation is measured directly, whereas in magnetometry it is inferred from the magnetic response. In our case, the frequency-dependent magnetic susceptibility is measured (between 10 Hz and 10 kHz) and used for computing heat loss.

Here, we present a coherent framework for the primary dissipation mechanisms relevant to ferrofluids and, more broadly, to magnetic nanoparticle systems. We present an interpretation of magnetometry and calorimetry measurements in the context of frequency-dependent magnetic dissipation, emphasizing the connections between different energy absorption processes and different measurement methods.

\section{The Emergence of Superparamagnetism}

Ferrofluids inherit their magnetic properties from the suspended magnetic nanoparticles. These nanoparticles are composed of magnetic materials, often certain iron or cobalt oxides. These magnetic properties can be traced back to the microscopic level, where they arise from spin ordering, primarily governed by exchange interactions. The simplest form of spin alignment is ferromagnetic ordering, in which neighboring spins align parallel to each other. However, in materials like magnetite (Fe$_3$O$_4$), we encounter ferrimagnetism, which, on the macroscopic level, is indistinguishable from ferromagnetism (although the detected magnetization is usually smaller) \cite{kittel,solyom1,torresdiaz2014}.

In bulk ferromagnetic materials, the presence of macroscopic dipole fields leads to the formation of a domain structure, which minimizes the magnetic energy. This domain configuration gives rise to memory effects in the magnetic response of the material. When an external magnetic field is applied, these materials retain a finite portion of their magnetization even after the field is removed; this is known as remanent magnetization. To regain a macroscopically demagnetized state, a finite, oppositely directed magnetic field (coercive field, $H_\mathrm{c}$) is required \cite{kittel,Review_ortega_pankhurst,garaio}.

\begin{figure}[ht]
    \centering
    \includegraphics[width=.6\linewidth]{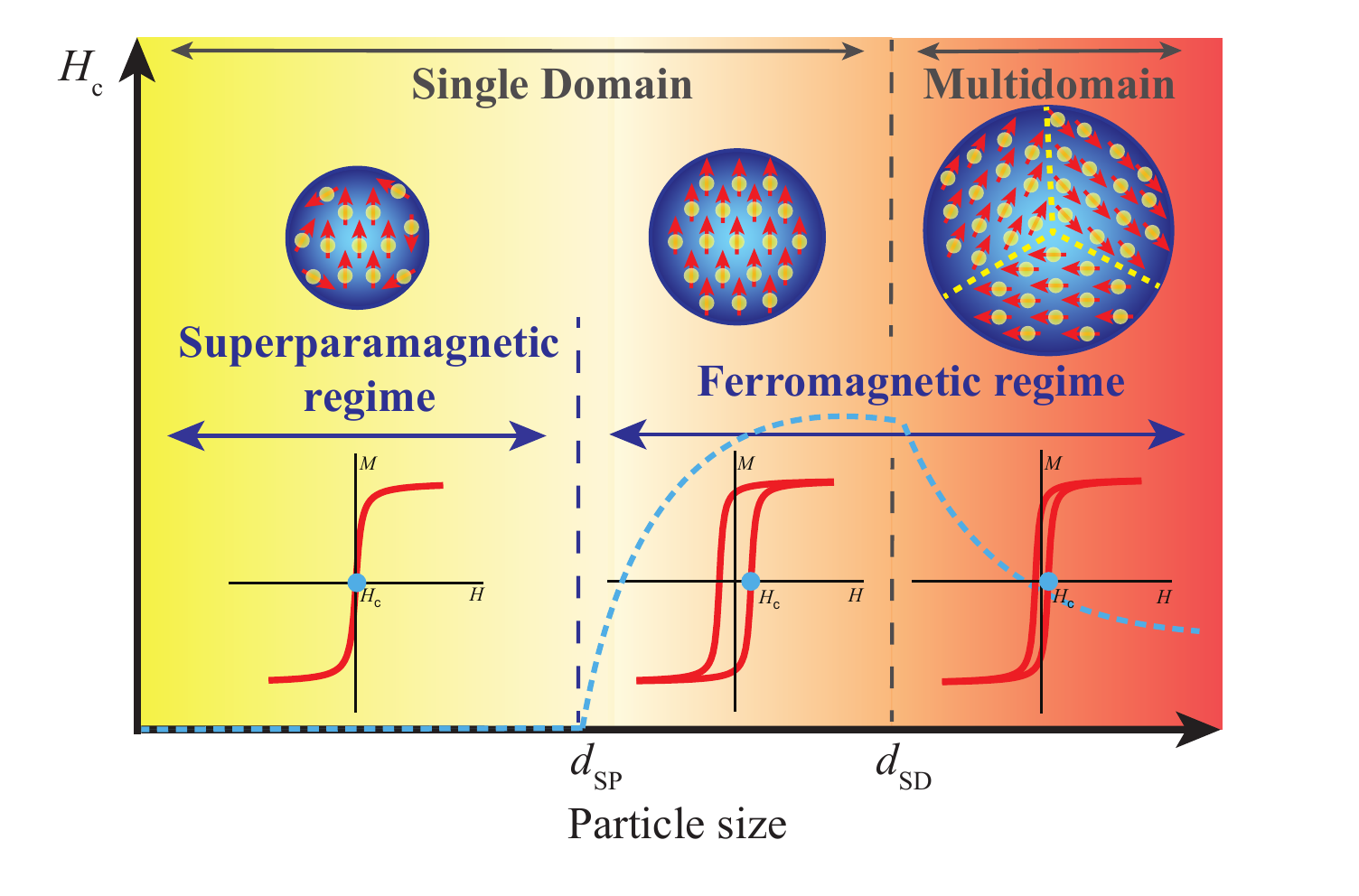}
    \caption{Qualitative illustration of the magnetic structure in ferromagnetic particles of varying sizes. The blue plot shows the dependence of the coercive field on sample diameter, alongside representative hysteresis curves corresponding to multi-domain, single-domain, and superparamagnetic regimes. From right to left: as the particle size decreases, a multidomain ferromagnetic particle becomes single-domain, with maximal coercive field. With further decreasing the particle size, thermal fluctuations are sufficient to allow a continuous change in the particle magnetization, which leads to superparamagnetism.}
    \label{fig:superpara}
\end{figure}

As Figure \ref{fig:superpara} suggests, when the size of ferromagnetic particles is decreased below a certain diameter threshold, typically on the order of $70-130$ nm for magnetite \cite{Butler1975, Baumgartner2013, Nel1955SomeTA, Leslie-Pelecky}, depending on material composition and geometry, they become single-domain (e.g., a macrospin). This critical particle size, $d_{\mathrm{SD}}$, is determined by the energy scales of the exchange interaction ($A$ exchange constant) and the magnetic anisotropy energy ($KV$, $K$ is the anisotropy constant) \cite{Butler1975, Aharoni2001, Nel1955SomeTA}:
\begin{equation}
    d_{\mathrm{SD}} \propto \left(\frac{A}{K}\right)^{1/2}.
\end{equation}

In this regime, the coercive field increases since magnetic reorientation is only possible by rotating the magnetization of the entire particle. This requires more energy than moving the domain walls, which means the reorientation of a few spins at the domain borders present in only multi-domain particles \cite{kittel, solyom1, Mederos-Henry2019}.

When the size of ferromagnetic particles is further reduced (for magnetite, usually in the range of $20-50$ nm \cite{Butler1975, Baumgartner2013}), the coercive field approaches zero, indicating the disappearance of magnetic memory effects and hysteresis in the quasi-static (DC) response. The critical particle diameter for this regime, $d_{\mathrm{SP}}$, is determined by scales of the thermal energy ($k_\mathrm{B}T$, $k_\mathrm{B}$ is the Boltzmann constant and $T$ is the temperature) and the magnetic anisotropy energy ($KV$) \cite{Butler1975, Leslie-Pelecky}:
\begin{equation}
    d_{\mathrm{SP}} \propto \left(\frac{k_\mathrm{B}T}{K}\right)^{1/3}.
\end{equation}

This regime is known as superparamagnetic, as the corresponding hysteresis curves resemble those of paramagnetic materials. However, the superparamagnetic response is much higher than that of regular paramagnets, as these particles consist of many strongly correlated atomic spins. As a result, they behave collectively as a single macrospin, with an effective spin quantum number on the order of $1{,}000$.

It is important to note that multi-domain samples exhibit zero net magnetization in the absence of an external field due to the random orientation of their internal domains. In contrast, single-domain particles possess a non-zero intrinsic magnetization, as the entire particle adopts a uniform magnetic orientation. Additionally, it is essential to recognize that superparamagnetism is not an intrinsic property of the material alone; it also depends on the measurement conditions, particularly the temperature and timescale. A particle may exhibit superparamagnetic behavior at one temperature or timescale, but appear \emph{blocked} under different conditions.

The investigated ferrofluid samples contain magnetite nanoparticles of a mean diameter of 10.6 nm. This material is known to exhibit ferrimagnetic ordering at sufficiently low temperatures, which is slightly different from ferromagnets on a microscopic level, but from a macroscopic point of view, the two are indistinguishable \cite{kittel}. Therefore, the properties discussed in this section apply to the investigated materials. 

For most technological applications, magnetic response and dissipation are desired to be well-controllable. This makes superparamagnetic materials advantageous, as they do not exhibit static magnetic memory effects. This is one of the features that make ferrofluids particularly suitable for a broad range of applications. In the liquid phase, the additional degrees of freedom associated with the particles moving in the carrier fluid promote superparamagnetic behavior even if the nanoparticles themselves show memory effects.

\section{Static and Dynamic Dissipation}

The most well-known dissipation mechanism in macroscopic ferromagnetic materials is static hysteresis. This mechanism arises from the domain wall motion induced by an external magnetic field. As the field is applied, the microscopically ordered regions (domains) aligned with the field direction grow, leading to an increase in the macroscopic magnetization of the sample. This process introduces memory effects in the alignment of microscopic magnetic moments and, consequently, in the overall magnetic state of the material. The energy required to move domain walls and the associated lag between magnetization and the applied field manifests in the magnetization-magnetic field diagrams as hysteresis, indicating dissipation in the magnetic system.

This memory effect is termed static hysteresis, as it is predominantly present in quasi-stationary magnetic fields reflecting the relatively large time scales involved in domain wall motion. It appears in the form of an open loop in the magnetization versus magnetic field curve (also known as a DC or static hysteresis loop). The energy dissipated per unit mass in a full magnetization cycle is proportional to the area enclosed by this loop as $E = -\oint \mathbf{M}\,\mathrm{d}\mathbf{B}$ \cite{Garaio_2014,Review_ortega_pankhurst}.

\begin{figure}[ht]
    \centering
    \includegraphics[width=\linewidth]{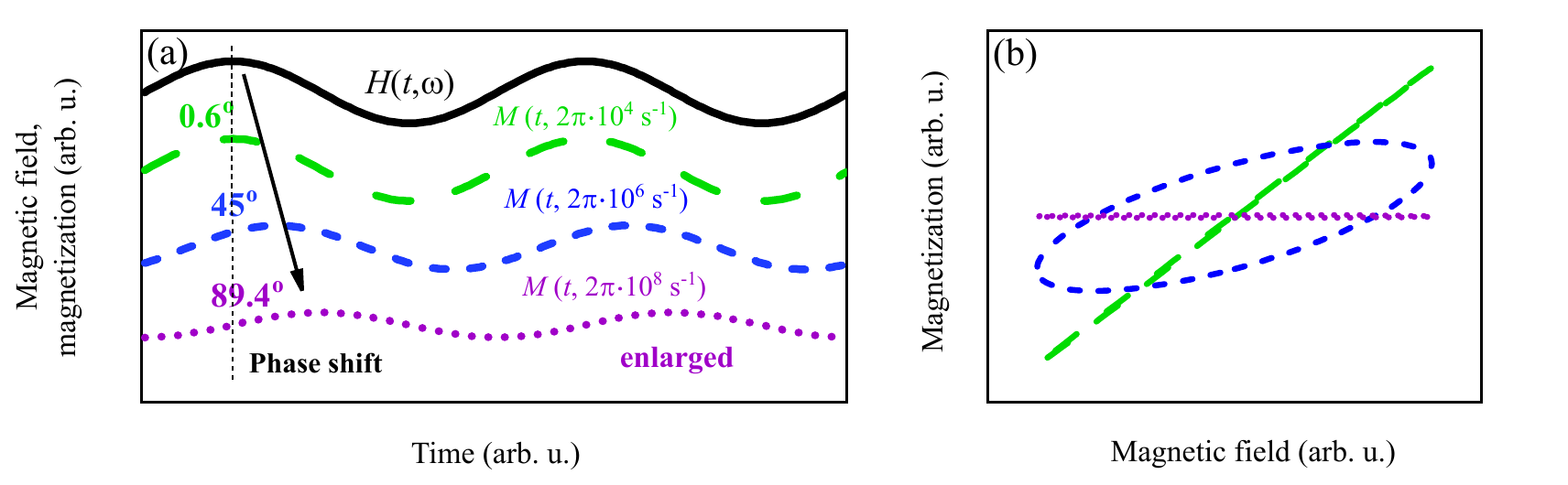}
    \caption{(a) Simulated time-dependence of magnetic field and magnetization at various frequencies, illustrating the frequency-dependent phase shift between excitation and response. (b) Magnetization as a function of magnetic field computed from the data shown in (a), highlighting the evolution of the dynamic hysteresis loop with frequency.}
    \label{fig:response}
\end{figure}

However, static hysteresis alone is not adequate to describe the magnetic response in alternating (AC) magnetic fields. In such cases, another mechanism -- the dynamic hysteresis -- has to be considered. Dynamic hysteresis can be described within the framework of linear response theory, which is valid for sufficiently weak excitations. Accordingly, the response functions of real physical systems are generally complex-valued, e.g., $\widetilde{\chi}\in\mathbb{C}$, implying a phase shift between the excitation and the response \cite{relax,torresdiaz2014,Review_ortega_pankhurst,Ivanov2023}. In magnetic systems, this corresponds to a phase difference between the external magnetic field $\mathbf{H}(\omega)$ and the resulting magnetization $\mathbf{M}(\omega)$. This phase shift depends on the frequency of the excitation (between zero for DC and $\pi/2$ for high-frequency fields), as shown in Figure \ref{fig:response}(a).

This phenomenon introduces memory effects into the system, as the momentary value of the magnetization depends not only on the current external field amplitude but also on its history. As a result, the relationship between magnetization and magnetic field becomes non-unique. The magnetic susceptibility, $\widetilde{\chi}(\omega)$, can be defined as the complex proportionality factor between the excitation field and the response:
\begin{equation}\label{equ:MH}
    \mathbf{M}(\omega) = \widetilde{\chi}(\omega) \mathbf{H}(\omega).
\end{equation}

In general, for linear materials, the susceptibility is a complex tensor. However, in the case of isotropic materials -- such as the ones considered here -- it reduces to a scalar quantity. The complex phase of the susceptibility is the phase shift between the excitation and the magnetic response \cite{solyom1}.

When the magnetization is plotted against the applied magnetic field, this phase shift leads to elliptical curves (neglecting magnetic saturation within the linear response approximation), as illustrated in Figure \ref{fig:response}(b). At very low and very high excitation frequencies compared to the characteristic frequency of the system, the ellipse collapses into a line with a slope corresponding to the DC susceptibility in the low-frequency limit and approaching zero in the high-frequency limit. For intermediate frequencies, the response forms a true ellipse with a finite enclosed area. This area corresponds to the energy dissipated (or the work done) during each excitation cycle.

Dynamic hysteresis is a universal feature of all magnetic systems, even those without a compound domain structure, which is necessary for static hysteresis. As such, static and dynamic hysteresis phenomena coexist in conventional ferromagnetic systems. In stark contrast, superparamagnetic materials display only dynamic hysteresis, due to the absence of domain wall motion.

The investigated sample demonstrates superparamagnetic behavior, as shown by the room-temperature DC hysteresis curve in Figure \ref{fig:DChist_T}. This property makes it suitable for studying purely dynamic dissipation mechanisms, as static memory effects are absent in the material. However, it can be observed that at sufficiently low temperatures, the nanoparticles lose their superparamagnetism and show memory effects and hysteresis even under quasi-static conditions. In the case of the investigated sample, this transition occurs below room temperature, as shown in Figure \ref{fig:DChist_T}.

\section{Magnetic Susceptibility Models}

\begin{figure}[ht]
    \centering
    \includegraphics[width=\linewidth]{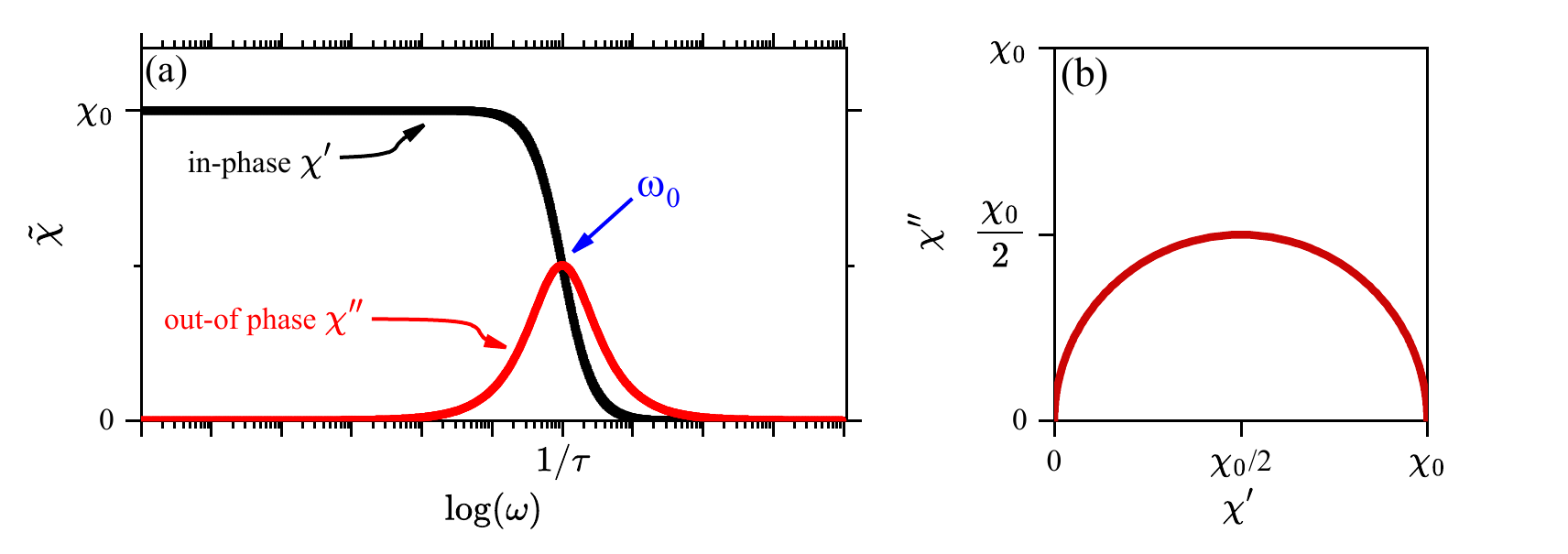}
    \caption{(a) Frequency dependence of the real and imaginary components of the magnetic susceptibility as predicted by the Debye relaxation model. (b) Cole--Cole plot of the Debye susceptibility model, showing the imaginary versus real parts of the susceptibility.}
    \label{fig:susc}
\end{figure}

The magnetic response of nanoparticle samples is often modeled using the relaxation-time approximation, which assumes that the time-dependent behavior of the magnetization is governed by a single characteristic relaxation time, denoted by $\tau$. The frequency dependence of such a response follows the so-called Debye model \cite{torresdiaz2014,Shliomis2002,Garaio_2014}, where the main features of the spectrum are illustrated in Figure \ref{fig:susc}(a):
\begin{equation}\label{equ:Debye}
    \widetilde{\chi}(\omega) = \frac{\chi_0}{1+\mathrm{i}\omega \tau} =\chi' - \mathrm{i}\chi''.  
\end{equation}
Here $\chi_0$ is the DC susceptibility, $\mathrm{i}$ is the imaginary unit, and $\omega$ is the angular frequency of the excitation. The parameter $\omega_0 = 1/\tau$ is the characteristic angular frequency of the system, which marks the peak in the imaginary component of the spectrum \cite{garaio,Review_ortega_pankhurst}.

The real and imaginary parts of this expression take the form (plotted for a specific $\omega_0$ value in Figure \ref{fig:susc}(a)):
\begin{equation}\label{equ:re_im}
    \begin{aligned}
         \chi' = \chi_0\frac{1}{1+\omega^2 \tau^2},\\
         \chi'' = \chi_0\frac{ \omega\tau}{1+\omega^2 \tau^2}.
    \end{aligned}
\end{equation}
The Cole--Cole plot of this function is shown in Figure \ref{fig:susc}(b). It takes the shape of a semicircle with a radius of $\chi_0/2$ and centered at $\chi' = \chi_0/2$, $\chi'' = 0$.

Real nanoparticle systems always have a finite size distribution, leading to deviations from the Debye model, because the characteristic frequency depends strongly on particle size \cite{Shokrollahi2013}. Furthermore, high-frequency contributions that are outside the accessible measurement range can further distort the observed susceptibility. To account for these effects, empirical fitting functions are commonly applied. These introduce empirical parameters, such as exponents $\alpha$ and $\beta$, and the high-frequency susceptibility $\chi_\infty$, modifying the standard Debye spectrum as follows \cite{Topping2018}:
\begin{equation}\label{equ:HN}
    \widetilde{\chi}(\omega)=\chi_\infty + \frac{\chi_0 - \chi_\infty}{\left(1+(\mathrm{i}\omega \tau)^\alpha\right)^\beta}.
\end{equation}

The expression described by Equation \eqref{equ:HN} is the so-called Havriliak--Negami model of susceptibility. Some special cases are often distinguished: when $\alpha = 1$, the model reduces to the Cole--Davidson form, and $\beta = 1$ recovers the generalized Debye model (or Cole--Cole model \cite{Holm2020101105}). When both $\alpha = 1$ and $\beta = 1$, the Debye model is recovered. For the exponents, $0<\alpha, \beta \leq 1$ reflect a physically relevant broadening of the relaxation spectrum \cite{Topping2018}. Such stretching represents a distribution of relaxation times, with components centered around different characteristic frequencies, consistent with systems exhibiting a non-trivial size distribution. The values $\alpha, \beta > 1$ have no physical meaning.

Nanoparticles in ferrofluids can become magnetically or mechanically immobilized at sufficiently low temperatures. Magnetic immobilization occurs when the system is cooled below the blocking temperature of the particles; in this range, the magnetization dynamics is suppressed. Mechanical immobilization, on the other hand, can be induced by freezing the carrier liquid, restricting the mechanical motion of the particles.

Magnetic relaxation arising from the mechanical rotation of the entire nanoparticle is referred to as Brownian relaxation, whereas the reorientation of the particle's magnetic moment without any mechanical rotation is known as N\'eel relaxation. The N\'eel relaxation time can be calculated as \cite{Leslie-Pelecky, Shokrollahi2013, garaio, Review_ortega_pankhurst, torresdiaz2014, Shliomis2002, Pucci2022}:
\begin{equation}
    \tau_\mathrm{N} = \tau_0~\mathrm{e}^{\frac{KV}{k_\mathrm{B}T}},
\end{equation}
where, $\tau_0$ is the material-specific attempt time.

The characteristic time of the Brownian relaxation is \cite{Shokrollahi2013, garaio, Review_ortega_pankhurst, torresdiaz2014, Shliomis2002, Pucci2022}:
\begin{equation}    
    \tau_\mathrm{B} = \frac{3\eta V_\mathrm{H}}{k_\mathrm{B}T},
\end{equation}
where $\eta$ is the dynamic viscosity of the carrier liquid and $V_\mathrm{H}$ is the mean hydrodynamic volume of the particles.

These two mechanisms typically operate on significantly different time scales. As a result, the magnetic response of a given ferrofluid is usually dominated by one of the two, allowing the particle dynamics to be described by a single effective relaxation time and treated in the formalism introduced in this section \cite{Shokrollahi2013, torresdiaz2014, Shliomis2002}.

\section{Dissipated Power in Different Susceptibility Models}

The dissipated energy per unit mass per excitation period ($P/m$, often denoted as $\mathrm{SAR}$ or Specific Absorption Rate \cite{Garaio_2014}) in a magnetic system is equal to the area of the hysteresis curve (magnetization-magnetic induction curve) \cite{kittel}: 
\begin{equation}\label{equ:dissipation}
    \frac{P(\omega)}{m} = -\frac1T\oint \mathbf{M}\,\mathrm{d}\mathbf{B}= -\frac{1}{T}\int_0^T \Re\left(\widetilde{\chi}(\omega) H_0\,\mathrm{e}^{-\mathrm{i}\omega t}\right)\cdot\frac{\mathrm{d}}{\mathrm{d}t}\left[\Re\left(\mu_0 \left[1+ \widetilde{\chi}(\omega)\right]H_0\,\mathrm{e}^{-\mathrm{i}\omega t}\right)\right]\,\mathrm{d}t,
\end{equation}
where $T$ is the excitation period, $\mu_0$ is the vacuum permeability, and a time-dependent excitation is assumed in the form $H(t) = \Re\{H_0 \mathrm{e}^{-\mathrm{i}\omega t}\}$, where $H_0$ is the amplitude and $\omega$ is the angular frequency. The validity of the linear response theory and the isotropy of the magnetic material are assumed. The mass of the sample is $m$, and $\widetilde{\chi}$ is understood to be susceptibility normalized to mass (e.g., mass susceptibility). All SAR values are normalized to the entire sample mass.

The result of the computation is proportional to the imaginary part of the dynamic magnetic susceptibility ($\chi''(\omega)$), and has the following form regardless of the details of the susceptibility spectrum:
\begin{equation}\label{equ:P}
    \mathrm{SAR}(\omega)=\frac{P(\omega)}{m} = \frac{1}{2} \mu_0 \omega H_0^2 \chi''(\omega).
\end{equation}

\begin{figure}[htb]
    \centering
    \includegraphics[width=\linewidth]{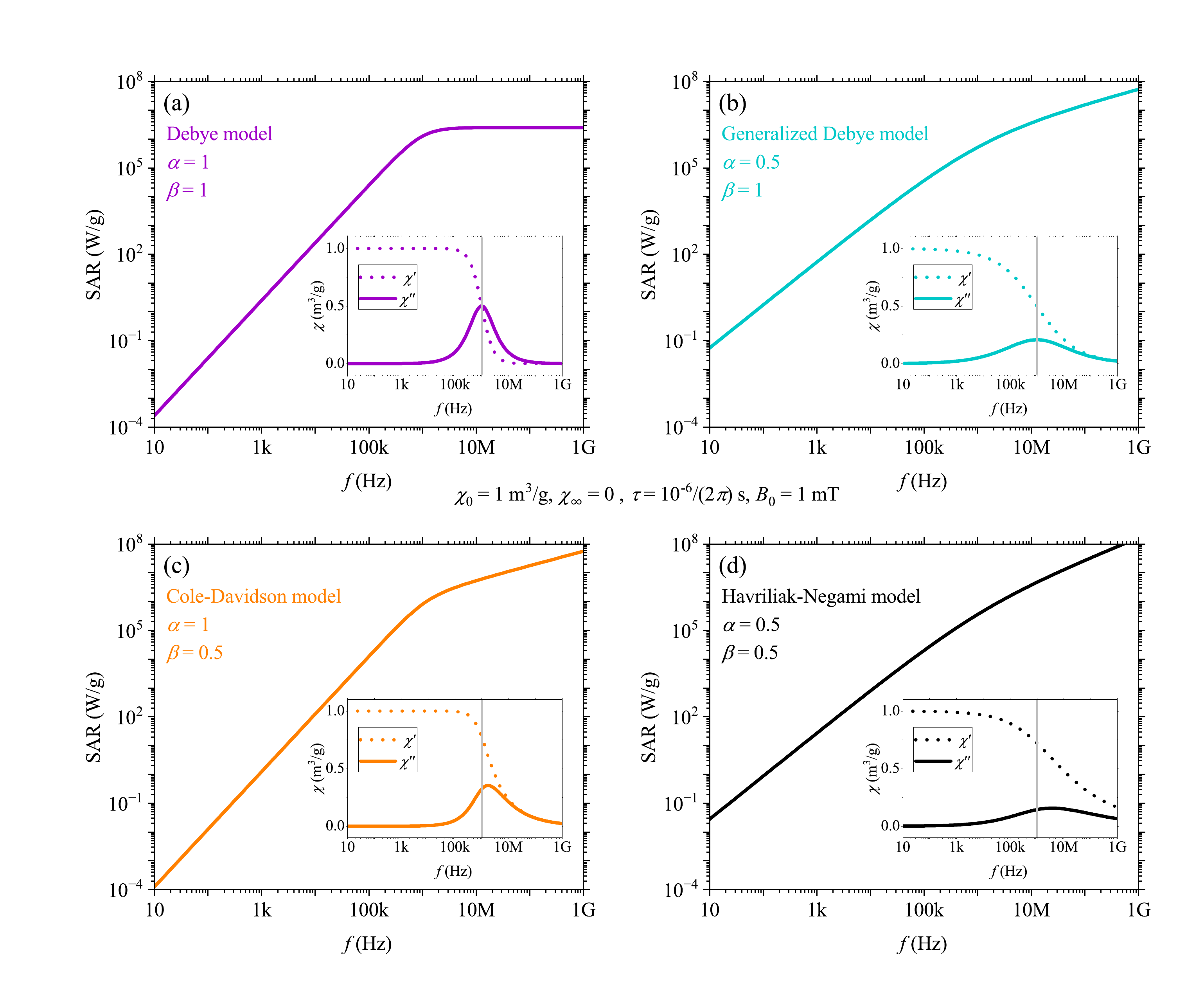}
    \caption{Computed SAR values as a function of frequency for (a) the Debye, (b) Generalized Debye, (c) Cole--Davidson, and (d) Havriliak--Negami susceptibility models. Computation parameters are denoted in the plots. Insets: corresponding real and imaginary parts of the magnetic susceptibility spectra.}
    \label{fig:SAR_calc}
\end{figure}

For the different susceptibility models described in the previous section, the dissipation versus frequency is displayed in Figure \ref{fig:SAR_calc}. For the Debye model (Figure \ref{fig:SAR_calc}(a)), the dissipated power is constant for high frequencies, whereas for the other models it is divergent, as the imaginary part of the susceptibility does not cut off fast enough to compensate for the factor of angular frequency appearing in Equation \eqref{equ:P}. In the framework of the Havriliak--Negami model, it is easy to obtain a condition for a convergent high-frequency limit of dissipated power: $\alpha\cdot\beta \geq 1$. Considering the physically relevant values of the parameters ($0<\alpha, \beta \leq 1$), this condition can be fulfilled only by the $\alpha=1=\beta$ case, namely the Debye model. It is important to note, however, that the empirical parameters introduced by the model often represent the effect of phenomena outside the measurement range, therefore, the model potentially does not provide a full description of the spectrum. A sign for this can be the not satisfying the Kramers--Kronig relations or the sum rule\cite{kittel,solyom1,Topping2018}.

It can be concluded that nanoparticles with a Havriliak-Negami susceptibility spectrum can be used in applications where the aim is to maximize the magnetic dissipation rate (e.g. in nanomagnetic hyperthermia), and it is optimal to use the highest possible frequencies at a certain excitation amplitude. This is limited by other considerations, such as instrumentation and physiological concerns, like the development of unwanted eddy currents at high frequencies \cite{HERGT}.

The real and imaginary parts of the magnetic susceptibility spectra corresponding to each model are shown as insets in Figure \ref{fig:SAR_calc}. As these demonstrate, decreasing $\alpha$ causes the peak in the imaginary part to widen, whereas decreasing $\beta$ causes this peak to shift to higher frequencies, broaden, and take an asymmetrical shape.

\section*{Materials and Methods}

We investigated the properties of a water-based ferrofluid containing spherical magnetite ($\mathrm{Fe_3 O_4}$) nanoparticles with a mean diameter of $10.6$ nm at a concentration of $30$ mg/mL. The nanoparticles were synthesized using a co-precipitation protocol described in Reference \onlinecite{Massart1981}. The fundamental magnetic properties of similar nanoparticles were previously characterized \cite{puerto, delaPresa2012,vajtai2024}.

Calorimetry is among the most direct methods for quantifying dissipation in a material. To perform these measurements, we employed a Fives Celes MP 6 kW induction heating system. The samples were placed within a solenoid and subjected to alternating magnetic fields at frequencies of $96$ kHz and $282$ kHz, with a field amplitude of $10$ mT. The time dependence of the sample temperature was monitored using a precision thermometer. From the recorded temperature changes, the dissipated power in the system was calculated.

In spite of their simplicity, calorimetry methods have certain limitations. These include relatively low precision due to uncontrolled thermal exchange with the environment, constraints in accessible excitation frequencies imposed by available circuit components, and the potential for inhomogeneous temperature distribution within the sample. Nevertheless, calorimetry remains a rapid and straightforward technique for quantifying dissipation and enables the real-time monitoring of thermal response.

To complement the calorimetric data, we also conducted magnetometry measurements, which offer higher precision in characterizing dissipation. Static hysteresis loops were recorded using a Quantum Design MPMS-5T system operating in DC SQUID mode. Measurements were performed at two representative temperatures: $5$ K and $300$ K.

Dynamic magnetic susceptibility spectra were recorded at $300$ K over a frequency range of $10$ Hz to $10$ kHz using a Quantum Design PPMS system operating in ACMS II mode. The amplitude of the applied AC magnetic field was $10$ Oe.

The measured susceptibility data were fitted using various models derived from the Havriliak--Negami spectrum (Equation \eqref{equ:HN}), including its special cases. The most precise fit was then extrapolated to higher frequencies. Subsequently, the susceptibility spectra were converted into specific absorption rate (SAR) values using Equation \eqref{equ:P}, enabling direct comparison with the calorimetric results despite the lack of overlap between the available frequency ranges of the two measurement methods. SAR values in the text are normalized to the entire sample mass. Measurement uncertainties were negligible across all datasets.

In our calculations, we converted magnetic units between SI and CGS systems; the conversion is described in the Appendix.

\section{Results and Discussion}

\begin{figure}[ht]
    \centering
    \includegraphics[width=0.6\linewidth]{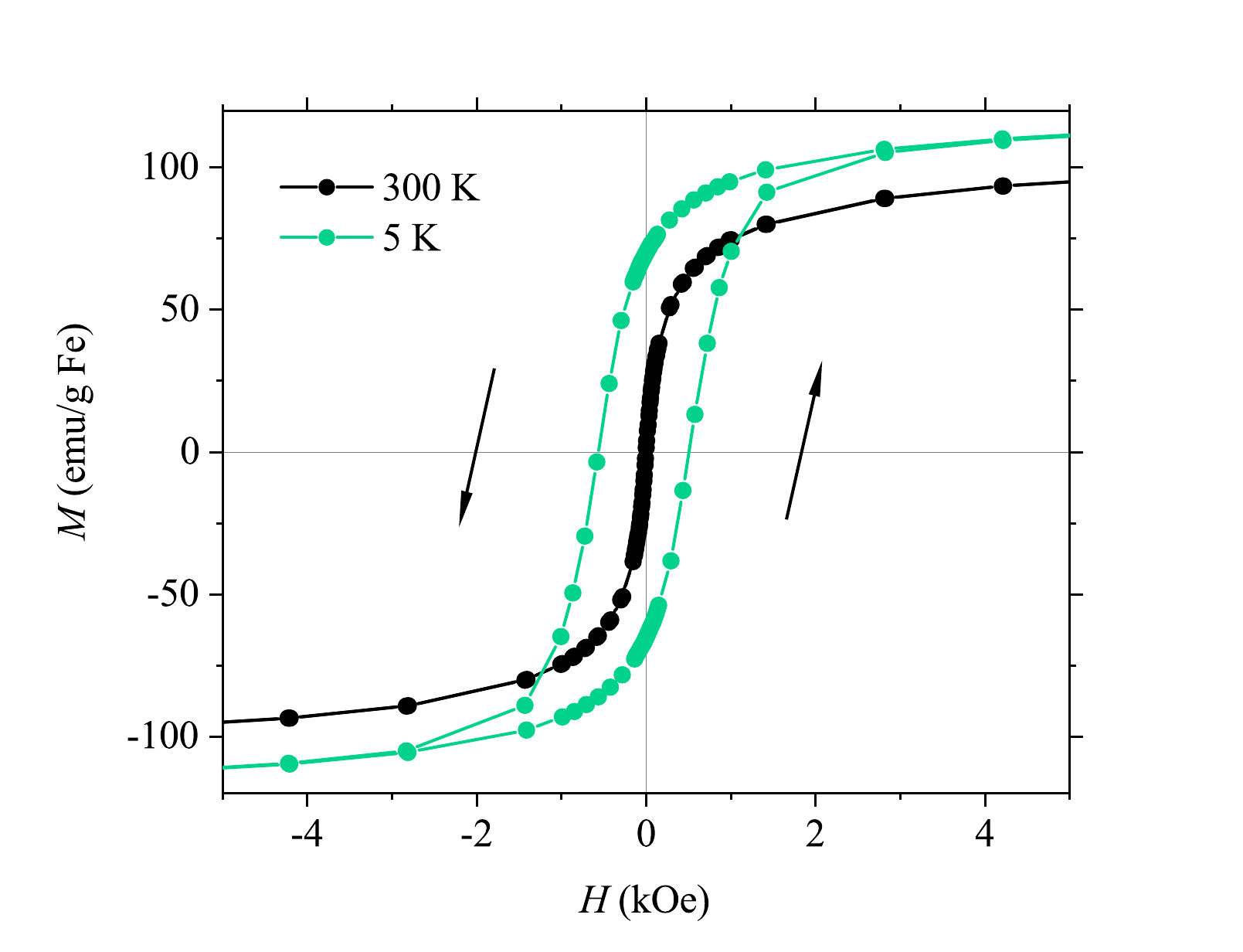}
    \caption{Measured static hysteresis curves at room and low temperature ($300$ K and $5$ K). Note the superparamagnetic nature of the sample at high temperature, and the appearance of the hysteresis loop at low temperature.}
    \label{fig:DChist_T}
\end{figure}

The superparamagnetic nature of the nanoparticles was confirmed by recording DC hysteresis curves at room temperature. As shown in Figure \ref{fig:DChist_T}, the magnetization data exhibit no memory effects, indicating the absence of static hysteresis. Consequently, the sample can be classified as superparamagnetic, and the only dissipation mechanism expected under these conditions is the dynamic (AC) dissipation.

Additionally, magnetic measurements demonstrate the immobilization of the magnetization at sufficiently low temperatures. The low-temperature DC hysteresis loop reveals clear memory effects and a nonzero loop area, signifying the presence of static hysteresis. It is also noteworthy that the saturation magnetization increases at lower temperatures, as reduced thermal agitation allows for more stable magnetic alignment.

The remainder of our measurements was conducted at room temperature. As previously demonstrated, the ferrofluid sample exhibits no static hysteresis under these conditions. Nevertheless, such superparamagnetic samples are widely used in applications that exploit energy dissipation via dynamic hysteresis. When subjected to an AC magnetic field, even superparamagnetic materials exhibit complex AC magnetic susceptibility, which is an indicator of dissipative behavior (see Equation \eqref{equ:P}).

\begin{figure}[ht]
    \includegraphics[width=\linewidth]{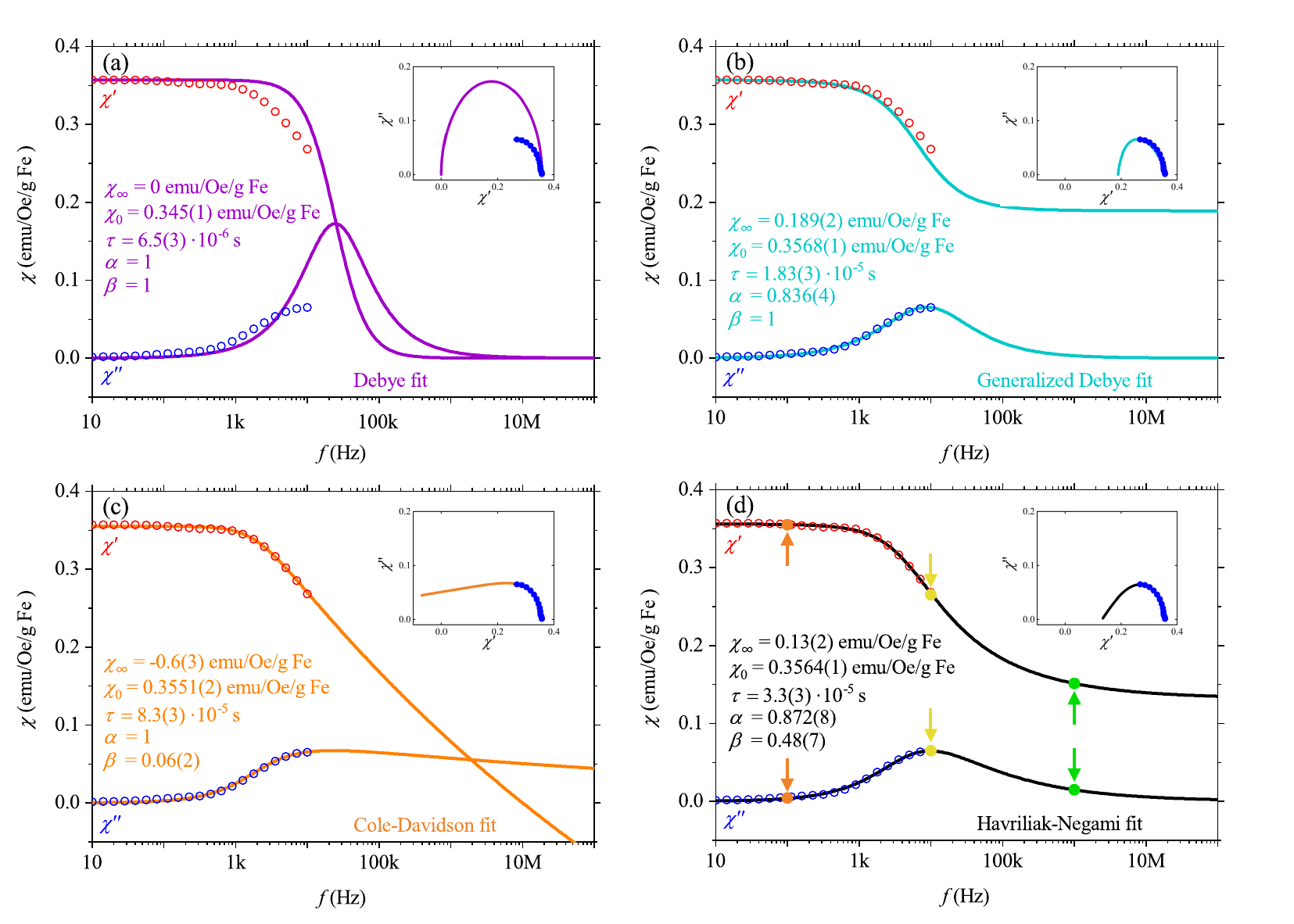} 
    \caption{Experimental data (open circles, same set of data in all subfigures) and possible models for fitting. (a) Debye model, (b) generalized Debye model, (c) Cole--Davidson model, (d) Havriliak--Negami model. Arrows and highlighted susceptibility values indicate data chosen for calculating magnetization-magnetic field curves in Figure \ref{fig:AChist_f}. Insets: Cole--Cole plot of the measured data and the individual fits.}
    \label{fig:fits}
\end{figure}

Figure \ref{fig:fits} presents the measured AC magnetic susceptibility spectrum at room temperature, along with several fitted models and their corresponding Cole--Cole plots. These fits and extrapolations are employed to approximate high-frequency behavior beyond the experimental measurement range, enabling a comparative analysis between magnetometry and calorimetry results, despite the absence of an overlapping frequency domain between the two techniques.

The simplest approach for modeling the frequency dependence of the susceptibility spectrum is the Debye model (Equation \eqref{equ:HN} with $\alpha = 1$ and $\beta = 1$, Figure \ref{fig:fits}(a)). Still, it fails to accurately capture the behavior of the measured data, yielding a relatively low adjusted $\mathrm{R}^2$ value (values for all the fitting functions are shown in Table \ref{tab:1}). This parameter reflects the accuracy of the fits, considering the number of fitting parameters as well to account for overparametrization. To illustrate the improvement offered by incorporating empirical parameters to account for unmeasured high-frequency effects, we considered alternative fitting models.

\begin{table}[!ht]
    \begin{center}
        \begin{tabular}{|c|c|c|c|c|}
        \hline
        Fitting model                   & Debye & Generalized Debye & Cole--Davison & Havriliak--Negami \\ \hline
        Adjusted $\mathrm{R}^2$ &    $0.99279$   &         $0.99996$         &    $0.99993$           &  $0.99997$                 \\ \hline
        \end{tabular}
    \caption{Adjusted $\mathrm{R}^2$ values for the different fitting functions. Note the increased values for the more complicated models despite the extra fitting parameters.}
    \label{tab:1}
    \end{center}
\end{table}

The generalized Debye model (Equation \eqref{equ:HN} with $\beta = 1$), depicted in Figure \ref{fig:fits}(b), significantly improves the fit, which is reflected in the adjusted $\mathrm{R}^2$ value as well. Nonetheless, noticeable deviations from the experimental data remain. A further enhancement is achieved using the Cole--Davidson model (Equation \eqref{equ:HN} with $\alpha = 1$), shown in Figure \ref{fig:fits}(c), which provides an excellent fit with an even higher adjusted $\mathrm{R}^2$. Despite the strong fit, the extrapolated high-frequency behavior predicts negative values for the real part of the susceptibility, which is unphysical for a superparamagnetic system.

\begin{figure}[ht]
    \centering
    \includegraphics[width=0.6\linewidth]{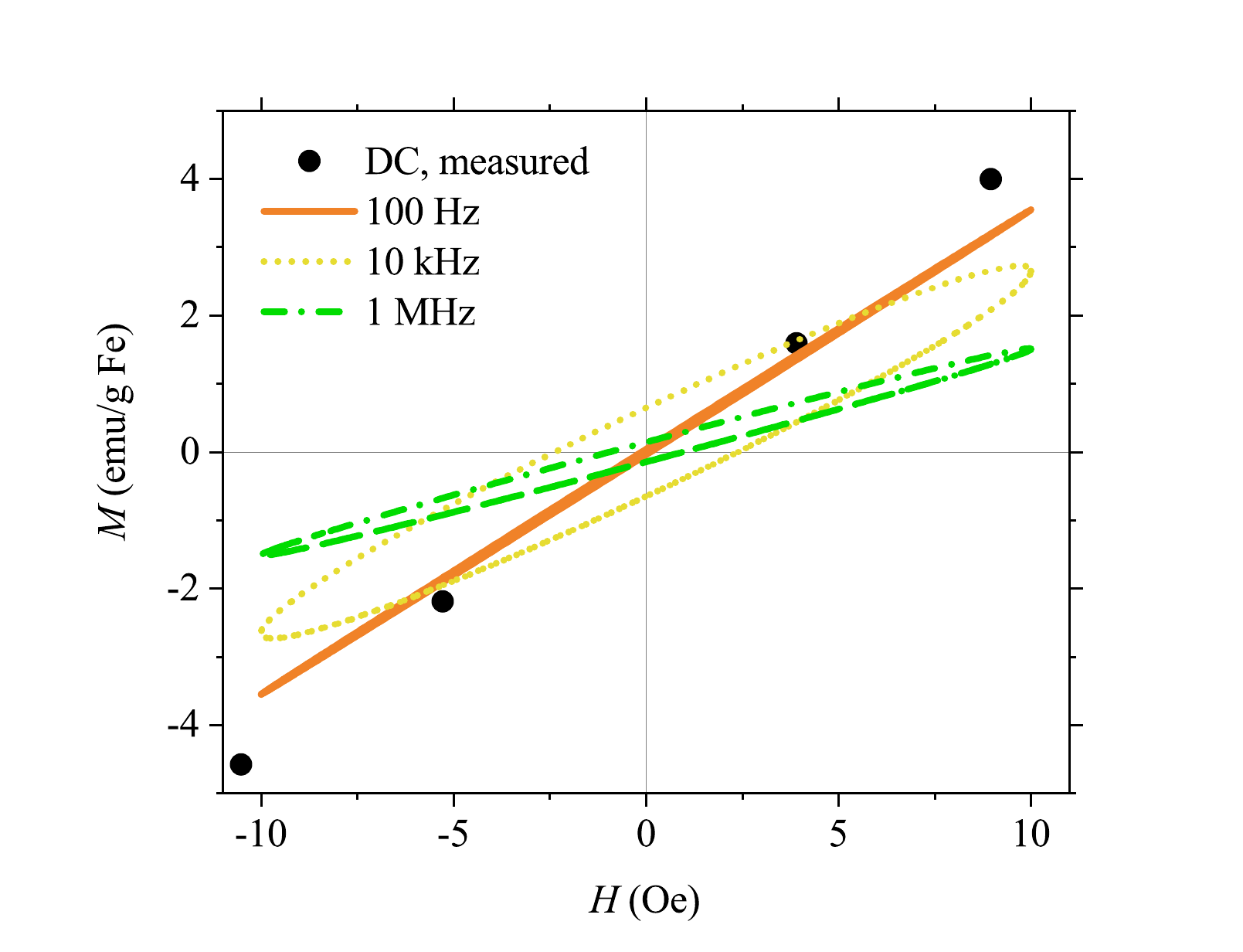}
    \caption{Calculated dynamic hysteresis curves at various frequencies and the measured static hysteresis data.}
    \label{fig:AChist_f}
\end{figure}

The best fit is given by the Havriliak--Negami model (Figure \ref{fig:fits}(d)). This model yields the highest adjusted $\mathrm{R}^2$ value, indicating an excellent fit. Therefore, this model is used to extrapolate towards higher frequencies in further calculations. In Figure \ref{fig:fits}(d), three specific frequency points are highlighted; these corresponding susceptibility values are later used to illustrate the relationship between static and dynamic hysteresis at various frequencies, demonstrating low-, high-, and intermediate-frequency characteristics of hysteresis curves. The comparison is shown in Figure \ref{fig:AChist_f}.

To demonstrate this, we calculated the time dependence of the magnetization response at the selected frequencies, assuming a sinusoidal external magnetic field (using Equation \eqref{equ:MH}). The resulting magnetization versus magnetic field plots, which are typically used to depict static hysteresis, exhibit open hysteresis loops even in this superparamagnetic system. These open loops indicate the presence of dissipation, arising solely from the finite frequency of the excitation. In this case, the dissipation originates not from static memory effects due to the domain reorientation but from the arising phase shift between the excitation and the response. This effect can be measured in real time as well \cite{Garaio_2014}.

The resulting dynamic hysteresis curves take the shape of ellipses, as the simulations performed under low-amplitude excitations are consistent with those used in AC susceptibility measurements, remaining within the linear response regime. At high excitation amplitudes, however, magnetic saturation becomes significant and must be taken into account in the calculations \cite{torresdiaz2014,Shliomis2002}.

At low frequencies, the curves appear as closed loops with a finite slope, reflecting a nonzero real part of the susceptibility and the absence of an imaginary component, indicating no dissipation. The calculated data align well with the results of the direct static magnetization measurements. The slight deviation is caused by the difference between the response to real quasi-static and low-frequency dynamic excitations.

At the highest frequency included in the calculations, the ellipsoidal loop collapses into a nearly flat line with zero slope, signaling that both the real and imaginary parts of the susceptibility approach zero. In contrast, the intermediate frequency displays an elliptical loop with a substantial area, revealing the presence of significant dissipation. This behavior corresponds to excitations around the characteristic frequency of the sample, where the imaginary part of the susceptibility, and thus energy loss, is maximal. This behavior mirrors the effect illustrated in Figure \ref{fig:response}(b).

\begin{figure}[ht]
    \includegraphics[width=\linewidth]{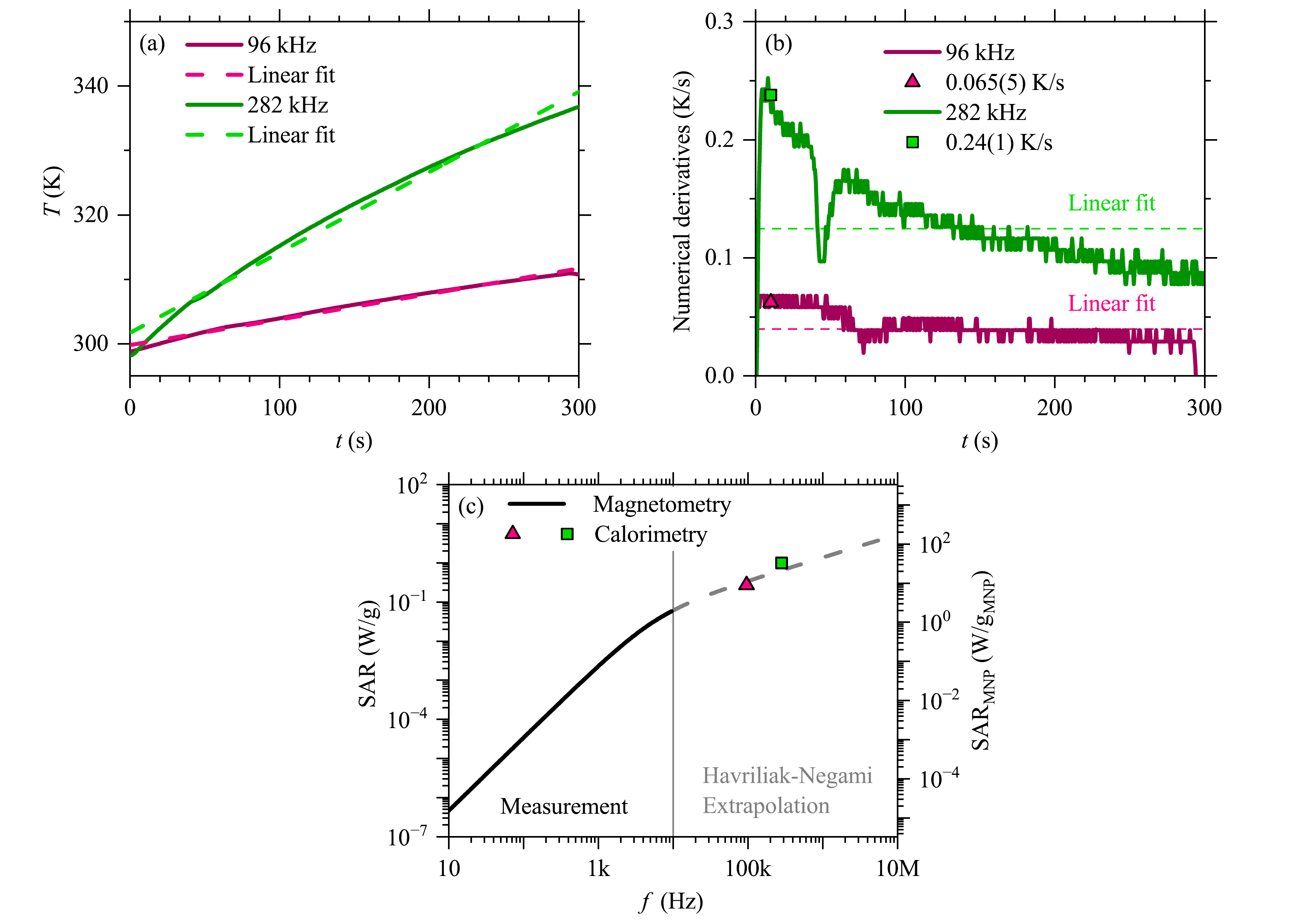} 
    \caption{(a) Measured temperature as a function of time during calorimetry experiments at different excitation frequencies, along with corresponding linear fits. (b) Numerical derivatives of the measured temperature data and linear fits, indicating heating rates. (c) Specific absorption rate (SAR) values, calculated from both AC magnetic susceptibility data and calorimetry measurements for comparison. Data are normalized to the whole sample mass. The axis on the right shows the scale normalized to the nanoparticle mass.}
    \label{fig:calorimetry}
\end{figure}

The susceptibility spectrum, extrapolated to high frequencies using the Havriliak--Negami model, was compared with the calorimetry measurements, as shown in Figure \ref{fig:calorimetry}.

In calorimetry measurements, the dissipated heat is expected to increase linearly under constant excitation power. Consequently, the sample temperature should rise linearly over time, and the dissipated power can be determined from the slope of a linear fit to the temperature–time data, as shown in Figure \ref{fig:calorimetry}(a). However, the linear fits clearly deviate from the measured data, which is even more apparent in Figure \ref{fig:calorimetry}(b), which displays the numerical derivatives of both the measurements and the fits.

This deviation arises from the thermal exchange between the sample and its environment. Since the heat current ($\mathbf{j}_Q$) is proportional to the negative temperature gradient ($-\mathbf{\nabla} T$):
\begin{equation}
    \mathbf{j}_Q = -k\mathbf{\nabla} T,
\end{equation}
according to Fourier's law of heat conduction \cite{liu1990fourier,solyom2}, where $k$ is the thermal conductivity of the medium. The increasing temperature of the sample leads to a growing temperature difference relative to the surroundings. This enhances the rate of heat loss, reducing the apparent net heating rate and causing the temperature rise to slow down over time. As a result, the measured temperature curve deviates from linearity, which also appears as the decreasing slope of the numerical derivative.

As a result, heat conduction has the greatest influence on the sample temperature when it rises significantly above room temperature. To more accurately assess the effect of magnetic heating separately from heat conduction, we substituted the slope of the fitted curve with the initial value of the numerical derivative ($\left.\frac{\mathrm{d}T}{\mathrm{d}t}\right|_{t=0}$) in our calculations, which corresponds to the moment when the sample is still approximately in thermal equilibrium with the environment, immediately after starting the heating process. The specific absorption rate (SAR) values were then calculated as follows:
\begin{equation}
    \mathrm{SAR} = \left.\frac{\mathrm{d}T}{\mathrm{d}t}\right|_{t=0}\cdot c,
\end{equation}
where the $c$ specific heat was assumed to be equal to the specific heat of water as the main constituent of the ferrofluid ($c~=~4.184~\mathrm{J/gK}$) \cite{ginnings1953heat}.

The SAR spectrum was also computed using the extrapolated magnetometry data shown in Figure \ref{fig:fits}(d), based on Equation \eqref{equ:P}. The susceptibility values were converted from CGS to SI units (as shown in the Materials and Methods section), and the normalization was adjusted from the iron content to the total sample mass.

In practice, we multiplied the imaginary part of the data shown in Figure \ref{fig:fits}(d) by $4\pi\times 10^{-3}$ to obtain susceptibility in SI units (see details in the Appendix). This was followed by a multiplication by the nanoparticle concentration ($C=30~\mathrm{\frac{kg}{m^3}}$) and a division by the density of water ($\rho=997~\mathrm{\frac{kg}{m^3}}$\cite{patterson1994measurement}, assuming the density being approximately equal to that of water as the main constituent of the ferrofluid) to obtain a mass susceptibility normalized to the whole sample mass instead of the iron content. We calculated the dissipated power per unit mass substituting to Equation \eqref{equ:P}. To summarize the entire calculation:
\begin{equation}
   \mathrm{SAR}(f)= \frac{P(f)}{m}=\frac{1}{2}\cdot \mu_0 H_0^2\cdot 2\pi f\cdot 4\pi\cdot 10^{-3} \cdot\chi_{\text{CGS}}''(f)\frac{C}{\rho},
\end{equation}
where $f$ is the irradiation frequency (note the change between arguments from angular frequency, $\omega = 2\pi f$, and frequency), and $H_0 = 10~\mathrm{Oe}=10^4/(4\pi)~\mathrm{\frac{A}{m}}$ is the amplitude of the time-dependent, sinusoidal excitation applied during AC susceptibility measurements. This SAR result is normalized to the entire sample mass, but it can be easily converted to values normalized to the nanoparticle mass contained in the ferrofluid.

A comparison of the magnetometry and calorimetry results is presented in Figure \ref{fig:calorimetry}(c). The dissipation rate per ferrofluid mass is shown on the left-hand-side axis ($\mathrm{SAR}$), whereas the right-hand-side axis shows the same SAR values normalized to the nanoparticle mass in the ferrofluid ($\mathrm{SAR_{MNP}}$). The magnetometry and calorimetry measurements yield consistent results. Minor discrepancies between them may arise from inaccuracies in the high-frequency extrapolation of the susceptibility spectrum, potentially due to nontrivial features beyond the measured frequency range. Nevertheless, the two measurement techniques appear to be reliable extensions of each other to form a unified, wider measurement range to determine the dynamic magnetic dissipation in ferrofluids.

\section{Conclusions}

We conducted DC and low-frequency AC magnetometry, along with calorimetry measurements, to investigate both static and dynamic hysteresis in a ferrofluid. Through detailed analysis, we showed that the dissipated power calculated from the magnetometry data agrees well with results from calorimetric measurements. These findings offer deeper insight into the magnetic behavior of nanoparticle systems and may support the future development and optimization of ferrofluids for specific applications. Based on our findings, magnetic and calorimetric methods might serve as valuable complementary techniques for determining dynamic magnetic dissipation in ferrofluids by extending the available frequency range.

\section*{Acknowledgments}
Work supported by the National Research, Development and Innovation Office of Hungary (NKFIH), and by the Ministry of Culture and Innovation Grants Number K137852, 149457, 2022-2.1.1-NL-2022-00004, TKP2021-NVA-02, and by the Ministerio de Ciencia e Innovacion, Spain, with Grants Number
TED2021-129254B-C21.

\bibliography{main.bbl}

\appendix
\section{Conversion of magnetic SI and CGS units}

As most commercial magnetometers use CGS units (by default), we briefly discuss the conversion to SI. A treatise on this issue can be found in, e.g., Reference \onlinecite{slichter1990}. The static magnetic moment data is returned in \textbf{emu} units, whereas the output of the ACMS is \textbf{emu/Oe}. Thus, the static data is divided by the excitation field amplitude in Oe units to proceed to susceptibility. In this context, \textbf{emu} values correspond to the total magnetic dipole moment ($\mu$) in the sample, whose unit is \textbf{A}$\cdot$\textbf{m}$^{\bm 2}$ (or \textbf{J}/\textbf{T}) in SI. We note that it would be more appropriate to use \textbf{erg}/\textbf{G} in CGS, which is a rare practice. The $\mu$ value in CGS is multiplied by $10^{-3}$ (this is due to the different Ampere definition in the two systems besides the apparent \textbf{cm} vs. \textbf{m} difference) to obtain the value in SI. Volume or mass susceptibility is obtained by dividing the \textbf{emu}/\textbf{Oe} values by the corresponding volume and mass in \textbf{cm}$^{\bm 3}$ or \textbf{kg}, respectively. Care has to be taken at this step for a heterogeneous sample (such as ferrofluids), as to what is meant by volume and sample mass. For consistency and comparison with literature results, we quote values divided by iron mass as the results of the dynamic susceptibility measurements. However, the proper quantity appearing in Maxwell's equations is the dimensionless volume susceptibility; thus, a normalization by the total sample volume is performed in analyzing our dissipation results.

We give mass susceptibility data in \textbf{emu}/\textbf{Oe}$\cdot$\textbf{g}, which is often (and inconsistently) quoted in the literature as units of \textbf{emu}/\textbf{g}. The volume and mass susceptibility in CGS is converted to SI by multiplying with $4\pi$ and $4\pi\times 10^{-3}$ to obtain dimensionless and \textbf{m}$^{\bm 3}$/\textbf{kg} units, respectively. The use of SI units is clearer where the magnetic moment dimensions, \textbf{Am}$^{\bm 2}$ is divided by the magnetic field (\textbf{A}/\textbf{m}) and the volume in \textbf{m}$^{\bm 3}$, resulting in dimensionless units.

\end{document}